\documentclass[reprint,prl]{revtex4-1}
\usepackage{amsmath,amssymb}
\usepackage[hypertex]{hyperref}

\def\section#1{{\it #1}.---}

\makeatletter

\def\half{{\frac{1}{2}}}

\def\unit{{1\kern-.65ex {\rm l}}}
\def\1{{1\kern-.65ex {\rm l}}}
\def\ap{\alpha'}
\def\ls{l_{\rm s}}
\def\gs{g_{\rm s}}

\def\CM{{\cal M}}
\def\CN{{\cal N}}

\def\bbR{{\mathbb{R}}}

\def\bbZ{{\mathbb{Z}}}

\def\dtt{\widetilde{dt}}

\makeatother

\begin{document}

\title{Exotic  branes and non-geometric backgrounds}
\date{\today}
\author{Jan de Boer and Masaki Shigemori}
\affiliation{Institute for Theoretical Physics, University of Amsterdam\\
 Valckenierstraat 65, 1018 XE Amsterdam, The Netherlands}
\preprint{ITFA10-07}
\begin{abstract}
When string/M-theory is compactified to lower dimensions, the
$U$-duality symmetry predicts so-called exotic branes whose higher
dimensional origin cannot be explained by the standard string/M-theory
branes.  We argue that exotic branes can be understood in higher
dimensions as non-geometric backgrounds or $U$-folds, and that they are
important for the physics of systems which originally contain no exotic
charges, since the supertube effect generically produces such exotic
charges.  We discuss the implications of exotic backgrounds for black
hole microstate (non-)geometries.
\end{abstract}
\maketitle


\section{Introduction}%
String theory includes various extended objects as collective
excitations, such as D-branes. The $U$-duality symmetry
\cite{Hull:1994ys} which maps these objects into one another has played
a pivotal role in the development of string theory and provided crucial
insights into its non-perturbative behavior.
When string/M-theory is compactified to lower dimensions, the
$U$-duality group gets enhanced, relating objects that were not related
in higher dimensions.  For example, M-theory compactified on $T^k$ has a
discrete $U$-duality group known as $E_{k(k)}(\bbZ)$ \cite{Hull:1994ys}.

In the lower ($d=11-k$) dimensional theory, if we start from a
codimension-two object obtained by partially wrapping a known 11D object
and act by $U$-duality on it, we start to produce objects whose
higher-dimensional origin is unknown; they are called \emph{exotic
branes} \cite{Elitzur:1997zn}. In type II language, some
of them have a tension proportional to $\gs^{-3}$ or $\gs^{-4}$.
For example, in type II string compactified on $T^2$, consider an
NS5-brane extending along six of the eight remaining noncompact
directions, not wrapping the internal $T^2$. If we perform a $T$-duality
along both $T^2$ directions, we obtain an exotic brane called $5^2_2$.
We will see later that this is a non-geometric background known as a
$T$-fold \cite{Hull:2004in}; as we go around the exotic brane, the
internal $T^2$ is nontrivially fibered and does not come back to itself,
but rather to a $T$-dual version.

One may think that such codimension-two objects are problematic due to
logarithmic divergences \cite{Greene:1989ya}, and that we do not need
them if we are concerned with the physics of non-exotic states.  However,
this is not true because of the supertube effect
\cite{Mateos:2001qs}---the spontaneous polarization phenomenon that
occurs when we bring a particular combination of charges together.  A
basic example is
\begin{align}
 \rm D0+F1(1)\to D2(1\psi)
 \label{supertube_D0+F1}
\end{align}
in which D0-branes and fundamental strings along $x^1$ polarize into a
D2-brane extending along $x^1$ and a closed curve in the transverse
directions parametrized by $\psi$.  Note that the D2 charge did not exist in
the original configuration.  Since the D2 is along a closed curve, there
is no net D2 charge, but only a D2 dipole charge.  The microscopic entropy
of the D0-F1 system can be recovered by counting the possible $\psi$ curves
that the system can polarize into \cite{Palmer:2004gu}.

Even if we start with a configuration of non-exotic charges, the
supertube effect can produce exotic charges.  Because the exotic charges
thus produced are dipole charges, there is no net exotic charge at
infinity and the problem of log divergences does not arise.  This implies
that exotic states are relevant even for the physics of systems which do
not originally contain exotic charges.

This is especially interesting in the context of black hole physics
where one typically considers a configuration of multiple (non-exotic)
charges.  We will argue later that the supertube effect and exotic
charges are relevant for the understanding of the physics of such black holes.

\section{Exotic states and their higher dimensional origin}%
If we compactify M-theory on $T^8$ or type IIA/B string theory on $T^7$
down to 3D, we obtain $\CN=16$ supergravity \cite{Marcus:1983hb} with
128 scalars (note that gauge fields can be dualized into scalars in
3D\@).  This theory has $E_{8(8)}$ as the $U$-duality group which is
broken to the discrete subgroup $E_{8(8)}(\bbZ)$ in string theory
\cite{Hull:1994ys}.  This $E_{8(8)}(\bbZ)$ is generated by $S$- and
$T$-dualities along the internal torus.  For example, in type IIB, a
D7-brane wrapped on the $T^7$ yields a point particle in 3D\@. Acting
with $S$- and $T$-dualities, we can obtain all other states in the
``particle multiplet'' of the $U$-duality group as explained in
\cite{Elitzur:1997zn}.

\begin{table}
  \begin{tabular}{|l|l|}
   \hline
   Type IIA&  
       P ({\bf 7}), F1 ({\bf 7}), D0 ({\bf 1}), D2 ({\bf 21}), D4 ({\bf 35}), D6 ({\bf 7}),\\
   & NS5 ({\bf 21}), KKM ({\bf 42}),
       $5_2^2$ ({\bf 21}), 
       $0^7_3$ ({\bf 1}), $2^5_3$ ({\bf 21}), \\
   & $4^2_3$ ({\bf 35}), $6^1_3$ ({\bf 7}), $0^{(1,6)}_4$ ({\bf 7}), $1^6_4$ ({\bf 7})\\
   \hline
Type IIB&
       P ({\bf 7}), F1 ({\bf 7}), D1 ({\bf 7}), D3 ({\bf 35}), D5 ({\bf 21}), D7 ({\bf 1}),\\
   &NS5 ({\bf 21}), KKM ({\bf 42}) ,
       $5_2^2$ ({\bf 21}), 
       $1^6_3$ ({\bf 7}), $3^4_3$ ({\bf 35}),\\
   &$5^2_3$ ({\bf 21}), $7_3$ ({\bf 1}), $0^{(1,6)}_4$ ({\bf 7}), $1^6_4$ ({\bf 7})\\
   \hline
   M-theory&  
   P ({\bf 8}), M2 ({\bf 28}), M5 ({\bf 56}), KKM ({\bf 56}),\\
   &
   $5^3$ ({\bf 56}), $2^6$ ({\bf 28}), $0^{(1,7)}$ ({\bf 8})\\
   \hline
  \end{tabular}
 \vspace*{-2ex}
 \caption{The point particle states and their multiplicities (boldface
 numbers) in string/M-theory compactified to 3D\@.}
 \label{table:exotic_states}
 \vspace*{-3ex}
\end{table}
In Table \ref{table:exotic_states}, we list the states in the particle
multiplet, including the exotic ones.
The notation for non-exotic states is standard; {\it e.g.}, P denotes a
gravitational wave and KKM denotes a Kaluza-Klein monopole.
For type II exotic states, we follow \cite{Elitzur:1997zn} and denote them
by how their mass depends on the $T^7$ radii.  The mass $M$ of $b^c_n$
depends linearly on $b$ radii and quadratically on $c$ radii.  For
$b^{(d,c)}_n$, $M$ also depends cubically on $d$ radii.  Moreover, $M$
is proportional to $\gs^{-n}$.  For example, the mass of $5^2_2$ 
depends on the radii $R_{i}$, $i=3,\dots,9$ of $T^7$ as $M=R_3\cdots
R_7(R_8 R_9)^2/\gs^2\ls^9$. We often display how the state ``wraps'' the
internal $T^7$ as $5^2_2(34567,89)$.  In this notation, the KK monopole
is denoted by $5_2^1$.  In M-theory, we use a similar notation except
that we drop the subscript $n$.  Using the transformation rules for the
radii $R_i$ and $\gs$ under $S$- and $T$-dualities, we can read off how
those states transform into one another \cite{Elitzur:1997zn}.

In the 3D theory, we would have 128 gauge fields if we could dualize all
the scalars into gauge fields \cite{Elitzur:1997zn}.  However, as we can
see from Table \ref{table:exotic_states}, there are as many as 240
charged particles \cite{Note1}, and this discrepancy ($240$ versus $128$) in the
3D theory is not understood \cite{Elitzur:1997zn}.  For
$d\ge 4$, this issue does not arise because we obtain just as many
charged particles as gauge fields \cite{Elitzur:1997zn}.
Here, we argue that the higher dimensional origin of exotic states
consists of non-geometric backgrounds or $U$-folds
\cite{Hellerman:2002ax, Hull:2004in}.

The argument is simple.  For example, consider a D7-brane wrapping
$T^7$, which is (magnetically) coupled to the RR 0-form $C^{(0)}$.  From
the 3D viewpoint, the D7-brane is a point particle and, as we go around it,
the 3D scalar $\phi=C^{(0)}$ shifts as
$\phi\to\phi+1$.  Namely, in 3D, the ``charge'' of the point particle is
nothing but the monodromy of the scalar $\phi$ around it.
This symmetry of shifting $\phi$ by one gets combined with other
dualities such as $S$- and $T$-dualities to form the $U$-duality group
$G(\bbZ)=E_{8(8)}(\bbZ)$, and the scalar $\phi$ gets combined with other scalars
into a matrix $M$ parametrizing the moduli space $\CM = SO(16)\backslash
E_{8(8)}(\bbR) / E_{8(8)}(\bbZ)$.  $U$-duality means that we can more
generally consider a 3D particle around which $M$ jumps by a general
$U$-duality transformation.
Thus, the ``charge'' of a 3D particle is defined by the $U$-duality
monodromy around it.  This can be regarded as a non-Abelian
generalization of the usual notion of $U(1)$ charges for which the
monodromy is an additive shift.  Clearly, the number of different
charges thus generalized is not in general equal to that of gauge
fields, which resolves the above puzzle.

If we lift such a monodromy to 10D/11D, we obtain a configuration in
which the internal space is nontrivially fibered as we go around the
particle and glued together by a $U$-duality transformation.  So, exotic
states correspond in 10D/11D to non-geometric backgrounds, or
``$U$-folds'' \cite{Hull:2004in}.  To our knowledge, the interpretation
of exotic states as $U$-folds has not appeared in the literature.  Note
that this construction differs from the more familiar $U$-folds in the
context of string compactifications \cite{Hellerman:2002ax}, where
$U$-duality is nontrivially fibered over a non-contractible circle in
the internal manifold, not over a contractible circle in the noncompact
directions.

Let us discuss how to classify ``charges'' defined by the monodromies
around them.  First, assume the existence of a charge with monodromy
$q$.  Namely, as we go around the particle in 3D, the moduli matrix $M$
undergoes the monodromy transformation $M\to Mq, q\in G(\bbZ)$.  If we
go to another $U$-duality frame by a $U$-duality transformation $U\in
G(\bbZ)$, then this becomes $\tilde M\to \tilde M \tilde q$ with $\tilde
M=MU$, $\tilde q=U^{-1}qU$.  So, in the dual frame, there exists a
charge with monodromy $\tilde q$.  Now, let us change the moduli $\tilde
M$ adiabatically to the original value $M$. If the charge is BPS, an
object with monodromy $\tilde q$ continues to exist, implying the
existence of the charge $\tilde q$ even for the original value of the
moduli $M$ (assuming that there is no line of marginal stability).  So,
starting from a charge $q$, we can generate other possible charges by
conjugation $\tilde q=U^{-1}qU$. Note that this does not mean that we
can generate all charges that exist in the theory by conjugation; there
can be many conjugacy classes in the group $G(\bbZ)$ and we cannot
generate charges in different conjugacy classes.  Also, there can be
non-BPS charges for which the above argument (of changing moduli
adiabatically) does not apply.

As a simple example, consider a D7-brane.  Around it, there is an
$SL(2,\bbZ)$ monodromy given by
$T=\left(\begin{smallmatrix}1&1\\0&1\end{smallmatrix}\right)$. Let us
conjugate this with a general $SL(2,\bbZ)$ matrix
$U=\left(\begin{smallmatrix}s&r\\ q&p\end{smallmatrix}\right)$.  The
conjugated charge is $\tilde T=U^{-1}TU=\bigl(\begin{smallmatrix}1+pq &
p^2\\ -q^2 & 1-pq\end{smallmatrix}\bigr)$, which is the monodromy of the
standard $(p,q)$ 7-brane.  Note that, although $U$ has 3 independent
parameters, the resulting charge $\tilde T$ has only 2 parameters. In
this sense, there exist only two different charges.

So, the set of all possible charges we can obtain from a given one $q$
by $U$-duality is its conjugation orbit.  This orbit is a subset of the
discrete non-Abelian ``lattice'' $G(\bbZ)$, and it makes no sense to ask
how many different charges there are in it.  However, to get a
qualitative idea of the size of the orbit, we can replace $G(\bbZ)$ by
the continuous group $G(\bbR)$ and count the dimension of the (now
continuous) orbit.
Using standard results on the dimensions of conjugacy classes in
non-compact groups and their relation to $sl(2)$-embeddings, one can
show that, {\it e.g.}\ for $G(\bbZ)=E_{8(8)}(\bbZ)$, the dimension of
the orbit generated by a 1/2-BPS object such as the D7-brane is 58
\cite{dBS}.  The 240 states in Table \ref{table:exotic_states} represent
240 particular points in this orbit, which can be obtained by
$U$-dualities preserving the rectangularity of the internal torus
\cite{Elitzur:1997zn}.

\section{Supergravity description of exotic states}%
To demonstrate the above idea, let us present the supergravity metric
for $5_2^2$ as an example.  This can be obtained by $T$-dualizing the KK
monopole metric transverse to its worldvolume.  The KK monopole
($5^1_2(56789,4)$) metric is
\begin{align}
\begin{split}
  ds^2&=dx_{056789}^2+H dx_{123}^2+H^{-1}(dx^4+\omega)^2,\\
 e^{2\Phi}&=1,\qquad d\omega=*_3 dH,\\
 H&=1+\textstyle\sum_p H_p,\quad 
 H_p={R_4/ (2|\vec x-\vec x_p|)},
\end{split}
\label{KKM_dilaton_omega}
\end{align}
where $\vec x_p$ are the positions of the centers in $\bbR^3_{123}$.
Now compactify $x^3$, which is the same as arraying centers at 
intervals of $2\pi \tilde R_3$ along $x^3$.  So,
\begin{align}
H&=1+ \sum_{n\in\bbZ}{R_4/[2(r^2+(x^3-2\pi \tilde R_3 n)^2)^{1/2}]}\notag\\
 &\approx 1+\sigma
 \log\bigl[{(\Lambda+\sqrt{r^2+\Lambda^2})/r}\bigr],\quad
  \label{KKM_harmonic}
\end{align}
where $\sigma={R_4/ 2\pi \tilde R_3}$ and we took a cylindrical
coordinate system $ds_{123}^2=dr^2+r^2d\theta^2+(dx^3)^2$. We
approximated the sum by an integral and introduced a cutoff $\Lambda$ to
make it convergent (see \cite{Sen:1994wr, Blau:1997du}).
$H$ in \eqref{KKM_harmonic} diverges as we send $\Lambda\to \infty$, but
this can be formally shifted away by introducing a ``renormalization
scale'' $\mu$ and writing
\begin{align}
 H(r)=h+\sigma\log({\mu/r})
 ~~\Rightarrow~~
 \omega = -\sigma\theta\, dx^3
\label{H=h+log}
\end{align}
where $h$ is a ``bare'' quantity which diverges in the
$\Lambda\to\infty$ limit.   The
log divergence of $H$ implies that such an infinitely long
codimension-two object is ill-defined by itself.  In physically sensible
configurations, this must be regularized either by taking a suitable
superposition of codimension-two objects \cite{Greene:1989ya} or, as we
will do later, by considering instead a configuration which is of higher
codimension at long distance.  

Now let us do a $T$-duality along $x^3$.  By the standard Buscher rule,
we obtain the metric for $5^2_2(56789,34)$:
\begin{align}
   ds^2&=H \,\, (dr^2+r^2d\theta^2) +HK^{-1}dx_{34}^2+dx_{056789}^2,
 \label{metric_522}\\
 B^{(2)}_{34}&=-K^{-1}{\theta \sigma},\quad 
 e^{2\Phi}=HK^{-1},\quad
 K\equiv H^2+\sigma^2\theta^2.\notag
\end{align}
In terms of the radii in this frame, $\sigma={R_3R_4/ 2\pi\ap}$.
Similar metrics of exotic states have been written down ({\it e.g.},
\cite{Blau:1997du} considered $6^1_3$), but they do not appear to have
been discussed in the context of $U$-folds.
As can be seen from \eqref{metric_522}, as we go around $r=0$ from
$\theta=0$ to $2\pi$, the size of the 3-4 torus does not come back to
itself:
\begin{align}
\begin{split}
  \theta=0   &:\quad G_{33}=G_{44}=H^{-1},\\
 \theta=2\pi&:\quad G_{33}=G_{44}={H/[H^2+(2\pi\sigma)^2]}.
\end{split}
 \label{T-monodromy_522}
\end{align}
This can be understood as a $T$-fold.  If we package the 3-4 part
of the metric and $B$-field in a $4\times 4$ matrix
\cite{Maharana:1992my}
\begin{align}
 M=\begin{pmatrix}G^{-1} & G^{-1}B \\ -BG^{-1} & G-BG^{-1}B \end{pmatrix}
\end{align}
then the $SO(2,2,\bbR)$ $T$-duality transformation matrix $\Omega$
satisfying $\Omega^t \eta \Omega=\eta,$ $ \eta=\left(\begin{smallmatrix}
0 & {\bf 1}_2 \\ {\bf 1}_2 & 0 \end{smallmatrix}\right),$ acts on $M$ as
$M\to M'=\Omega^t M\Omega$\@.  It is easy to see that the matrix
\begin{math}
 \Omega=
 \left(\begin{smallmatrix} {\bf 1}_2 &  0 \\ 2\pi \sigma & {\bf 1}_2 \end{smallmatrix}\right)
\end{math}
relates the $\theta=0,2\pi$ configurations in \eqref{T-monodromy_522}.
Namely, $5^2_2$ is a non-geometric $T$-fold with the monodromy $\Omega$.

Although the mass of such a codimension-two object is not strictly
well-defined, we can still compute it by the following {\it ad hoc}
procedure.  The Einstein metric in 3D is given by $ds_3^2=-dt^2+H
dx^2_{12}$.  If $\gamma_{ij}$ is the spatial metric for constant $t$
slices and $G_{\mu\nu}$ is the Einstein tensor, we can compute
$\sqrt{\gamma}\,G_0^0 ={1\over 2}\partial_i^2\log H$.  So, the energy is
\begin{align*}
 M&
 =-{1\over 8\pi G_3}\int d^2x\sqrt{\gamma}\,G_0^0
 =-{1\over 16\pi G_3}\int dS\cdot \nabla\log H.
\end{align*}
If we use \eqref{H=h+log} and assume that
$H(r=\infty)=1$, then
\begin{align*}
 M
 ={1\over 16\pi G_3}\left.\left[ {{2\pi \sigma}\over H(r)}\right]\right|_{r\to \infty}
 ={(R_3 R_4)^2 R_5\cdots R_9 \over \gs^2 \ls^9},
\end{align*}
as expected of a $5^2_2(56789,34)$.  Here, we used $16\pi
G_3={\gs^2\ls^8/ R_3\cdots R_{9}}$. Although the $5^2_2$ changes the
asymptotics, setting $H(r=\infty)=1$ effectively puts it in an
asymptotically flat space and allows us to compute its mass.

Similarly, one can derive the metric for other exotic states appearing
in Table \ref{table:exotic_states}\@.  The metric provides an
approximate description, just as for ordinary branes, unless the tension
of the exotic branes is proportional to $g_s^{-3}$ or $g_s^{-4}$ and the
metric description breaks down.


\section{Supertube effect and exotic states}%
The above exotic $5^2_2$ brane appears in $d=3$ dimensions, but exotic
states are relevant to physics in $d\ge 4$ dimensions as well.  By
dualizing the basic supertube effect \eqref{supertube_D0+F1}, we can
derive the following spontaneous polarization:
\begin{align}
 \rm D4(6789)+D4(4589)\to 5^2_2(4567\psi,89).
 \label{D4D4puffup}
\end{align}
The configuration on the left can be thought of as a configuration in
4D, which puffs up into an extended configuration of an exotic dipole
charge along a curve in $\bbR^3_{123}$.  Such exotic \emph{dipole}
charges do not change the asymptotics of spacetime.  Note that the
original configuration of D4-branes is part of the standard D0-D4
configuration used for the black hole microstate counting in 4D
\cite{Maldacena:1997de}.  So, to understand the physics
of such black holes, it is unavoidable to consider exotic charges.

The supergravity solution for the configuration \eqref{D4D4puffup} can
be obtained by dualizing the solution for the supertube
\cite{Emparan:2001ux} and is given by
\begin{align}
  ds^2&=
 -f_1^{-\half}f_2^{-\half}(dt-A)^2
 +f_1^{\half}f_2^{\half}dx_{123}^2
 +f_1^{\half}f_2^{-\half}dx_{45}^2\notag\\
 &\qquad +f_1^{-\half}f_2^{\half}dx_{67}^2
 +{f_1^{\half}f_2^{\half} h^{-1}}dx_{89}^2,\\
 e^{2\Phi}&={f_1^{\half}f_2^{\half} h^{-1}},\quad
 B^{(2)}_{89}={\gamma h^{-1}},\quad
 C^{(3)}=-\gamma \rho +\sigma,
 \notag
\end{align}
where $h=f_1 f_2+\gamma^2$ and $\rho,\sigma$ are 3-forms given by
\begin{align*}
\begin{split}
 \rho&= (f_2^{-1}\dtt-dt)\wedge dx^4\wedge dx^5
 + (f_1^{-1}\dtt-dt)\wedge dx^6\wedge dx^7\\
 \sigma&= (\beta_1-\gamma dt)\wedge dx^4\wedge dx^5
       +(\beta_2-\gamma dt)\wedge dx^6\wedge dx^7.
\end{split}
\end{align*}
with $\dtt=dt-A$.  The $\psi$ curve in \eqref{D4D4puffup} is an
arbitrary closed curve in $\bbR^3_{123}$, and $f_{i=1,2},A$ are harmonic
functions sourced along the curve \cite{Emparan:2001ux}; see {\it e.g.}\
\cite{Dabholkar:2006za} for their explicit expressions.  The 1-form
$\beta_{i}$ and scalar $\gamma$ are related to $f_i,A$ by $d\beta_i=*_3
df_i, d\gamma=*_3 dA$.  In particular, for a circular curve,
they can be explicitly written down \cite{Emparan:2001ux,
Dabholkar:2006za}, including $\gamma,\beta_i$ \cite{dBS}.
As one goes around the curve, $\gamma$ undergoes a shift $\gamma\to
\gamma+q$ with $q$ a constant proportional to the $5^2_2$ dipole charge.
This gives rise to a monodromic structure in the metric and the $B$-field,
similar to the one in \eqref{metric_522}.  Because the exotic $5^2_2$
charge in \eqref{D4D4puffup} is merely a dipole charge, the 4D spacetime
is still asymptotically flat.

The D0-D4 system studied in the context of 4D black hole microstate
counting \cite{Maldacena:1997de} involves more stacks
than \eqref{D4D4puffup}: D0, D4(6789), D4(4589), D4(4567).  If we bring
these four stacks together, each pair is expected to undergo the
supertube effect:
\begin{align}
 {\rm D0}~~
 \begin{array}{l}
  \rm D4(6789)\\[-.5ex]
  \rm D4(4589)\\[-.5ex]
  \rm D4(4567)
 \end{array}
 ~
 \to
 ~
 \begin{array}{l}
  \rm NS5(6789\psi)\\[-.5ex]
  \rm NS5(4589\psi)\\[-.5ex]
  \rm NS5(4567\psi)
 \end{array}
 \begin{array}{l}
  \rm 5^2_2(6789,45\psi)\\[-.5ex]
  \rm 5^2_2(4589,67\psi)\\[-.5ex]
  \rm 5^2_2(4567,89\psi)
  \end{array}
 \label{D0D4D4D4puffup}
\end{align}
However, the charges on the right of \eqref{D0D4D4D4puffup} include
combinations of charges which can puff up again.  {\it A priori}, there
is no reason to exclude such further puff-ups which will produce all
kinds of exotic charges appearing in Table \ref{table:exotic_states},
assuming that such puff-ups do not break supersymmetry. As a different
example, take the 3-charge M2 system \cite{Bena:2004de} which is a well
studied configuration in the context of 5D black hole microstate
counting \cite{Strominger:1996sh}.  In this case, even if we restrict to
codimension-two puff-ups, the following sequence seems logically
possible:
\begin{align*}
 \begin{array}{l}  
  \rm M2(56)\\[-.5ex]
  \rm M2(78)\\[-.5ex]
  \rm M2(9A)
 \end{array}
 \to
 \begin{array}{l}
  \rm M5(\psi 789A)\\[-.5ex]
  \rm M5(\psi 569A)\\[-.5ex]
  \rm M5(\psi 5678)
 \end{array}
 \to
 \begin{array}{l} 
  \rm 5^3(\phi 789A,\psi 56)\\[-.5ex]
  \rm 5^3(\phi 569A,\psi 78)\\[-.5ex]
  \rm 5^3(\phi 5678,\psi 9A)
 \end{array}
 \to\cdots
 \label{puffup:M2M2M2}
\end{align*}
where
``A'' denotes the $x^{10}$ direction.  Namely, the system can polarize
into exotic branes extended along a 2-dimensional surface parametrized
by $\psi,\phi$ in $\bbR^{4}_{1234}$.
In the 2-charge system \cite{Lunin:2001jy}, entropy comes from the Higgs
branch of the worldvolume theory associated with the intersection of two
stacks of branes.  In gravity, the same entropy is explained by the
degrees of freedom coming from the fluctuations of the one-dimensional
geometric object which is the result of puffing up the intersection
\cite{Rychkov:2005ji}.  In the 3-charge system, the triple intersection
of three stacks of branes leads to a more complicated Higgs branch and
larger entropy. It is conceivable that the fluctuations of the above
2-dimensional exotic object that naturally appears, with its larger
number of degrees of freedom, account for the entropy of the 3-charge
system.
It would hence be very interesting to construct non-geometric solutions
involving such exotic charges to see if they can really reproduce the
expected entropy.
The fact that the 3-charge supergravity microstates constructed thus far
(see {\it e.g.}\ \cite{Bena:2005va, deBoer:2009un}) do
not seem enough to account for the entropy of the 3-charge black hole
\cite{deBoer:2009un} may be related to the non-geometric nature of
exotic branes that have been overlooked.


%
%
%

\begin{acknowledgments}
We thank I.~Bena, J.~Maldacena, P.~McFadden, N.~Obers, K.~Papadodimas,
M.~Rangamani, J.~Raeymaekers, T.~Takayanagi, and P.~West for
discussions.  This work was supported by the Foundation of Fundamental
Research on Matter (FOM) and by an NWO Spinoza grant.
\end{acknowledgments}



\end{document}